# Color Differences between Clockwise and Counterclockwise Spiral Galaxies

**Lior Shamir** [1,*]

[1] Lawrence Technological University, Michigan, MI 48075, USA; E-Mails: lshamir@mtu.edu (L.S.)

\* Author to whom correspondence should be addressed; E-Mail: lshamir@mtu.edu (L.S.);
Tel.: +1-248-204-3512; Fax: +1-248-204-3518.

**Abstract:** While spiral galaxies observed from Earth clearly seem to spin in different directions, little is yet known about other differences between galaxies that spin clockwise and galaxies that spin counterclockwise. Here we compared the color of 64,399 spiral galaxies that spin clockwise to 63,215 spiral galaxies that spin counterclockwise. The results show that clockwise galaxies tend to be bluer than galaxies that spin counterclockwise. The probability that the color differences can be attributed to chance is ~0.019.

**Keywords:** Spiral galaxies; galaxy spin; star forming galaxies

## 1. Introduction

The physics of the rotation of spiral galaxies is not fully understood [1], and previous studies show disagreement between the luminous and dynamical mass of spiral galaxies [2]. While current proposed explanations include dark halos [3] or modifications of the Newtonian dynamics [4,5], the physics of the rotation of spiral galaxies is not yet fully understood.

An obvious morphological difference between spiral galaxies is the handedness - some spiral galaxies spin clockwise, while other spiral galaxies rotate counterclockwise. However, although the difference in spin direction is highly visible, it is not clear whether the handedness is linked to other characteristics of the galaxies. A large-scale correlation between the galaxy handedness and other characteristics can indicate on asymmetry at a cosmological scale.

Here we analyze 127,614 spiral galaxies imaged by Sloan Digital Sky Survey (SDSS) to investigate whether there are differences between the color of spiral galaxies that spin clockwise and the color of spiral galaxies that spin counterclockwise.

**2. Data**

The dataset included galaxies imaged by SDSS [6], and classified as spiral galaxies by Galaxy Zoo [7]. Then, the galaxies were analyzed using the Ganalyzer method [8,9], and were separated as described in [10] to 64,399 spiral galaxies that spin clockwise and 63,215 spiral galaxies that spin counterclockwise. The Ganalyzer algorithm that was used to classify between clockwise and counterclockwise galaxies is described thoroughly in [8,10,12].

The declination of the galaxies in the dataset was between ~11.2$^o$ to ~70.3$^o$, and the redshift of the galaxies was smaller than 0.3. Some RA ranges had very few or no galaxies in them. The RA range 60$^o$-90$^o$ had less than 100 galaxies, and the range 270$^o$-300$^o$ had no galaxies in the dataset.

Table 1 shows the number of galaxies in each RA range.

**Table 1.** The number of galaxies in each RA range.

| RA range (degrees) | Number of galaxies |
|---|---|
| 0-30 | 3682 |
| 30-60 | 1959 |
| 60-90 | 71 |
| 90-120 | 3111 |
| 120-150 | 21574 |
| 150-180 | 27401 |
| 180-210 | 26938 |
| 210-240 | 27291 |
| 240-270 | 10235 |
| 270-300 | 0 |
| 300-330 | 2023 |
| 330-360 | 3329 |

The magnitudes used in this study are the cmodel magnitude values taken from the PhotObjAll table ("u", "g", "r" ,"i", "z" fields) of SDSS DR7. The cmodel magnitude is defined by

$$cmodel = fracDeV \cdot F_{deV} + (1 - fracDeV)F_{exp},$$

where *fracDeV* is the coefficient of the de Vaucouleurs term, and $F_{deV}$ and $F_{exp}$ are the de Vaucouleurs and exponential fluxes of the celestial object.

## 3. Results

Table 2 shows the mean and standard error of the mean of the u-g, g-r, r-i, i-z of the galaxies that rotate clockwise and galaxies that rotate counterclockwise. For each color the table also shows the two-tailed probability P that the two means are not different. The probability P that the two means appear to be different due to mere chance was computed using the unpaired t-test with different sample sizes, which given the means, sample sizes, and standard error of the means (or standard deviation) determines the statistical significance P that the two means are not different.

**Table 2.** Mean and standard error of u-g, g-r, r-i, i-z of clockwise and counterclockwise spiral galaxies. *P* is determined by two-tailed unpaired t-test.

| Color | Clockwise | Counterclockwise | P |
|---|---|---|---|
| u-g | ~1.51831±0.001701 | ~1.52405±0.00176 | ~0.0194 |
| g-r | ~0.77590±0.001193 | ~0.77772±0.00117 | ~0.2733 |
| r-i | ~0.39161±0.000798 | ~0.39285±0.00076 | ~0.2606 |
| i-z | ~0.27045±0.000541 | ~0.27012±0.00065 | ~0.6886 |

As the table shows, most of the differences between the means are not statistically significant except for u-g, in which the two-tailed probability that the two means are not different is ~0.0194.

Figure 1 shows the means and standard error of u-g, g-r, r-i, i-z in different $30^o$ RA slices. As can be learned from the figure, the u-g is higher in galaxies that spin counterclockwise in almost all RA ranges. The most significant two-tailed t-test of the difference between the mean is in the RA range $180^o$-$210^o$, and the probability that the difference between the means can be attributed to mere chance is ~0.0211.

**Figure 1.** Mean and standard error of u-g, g-r, r-i, i-z of spiral galaxies with different handedness in different RA ranges.

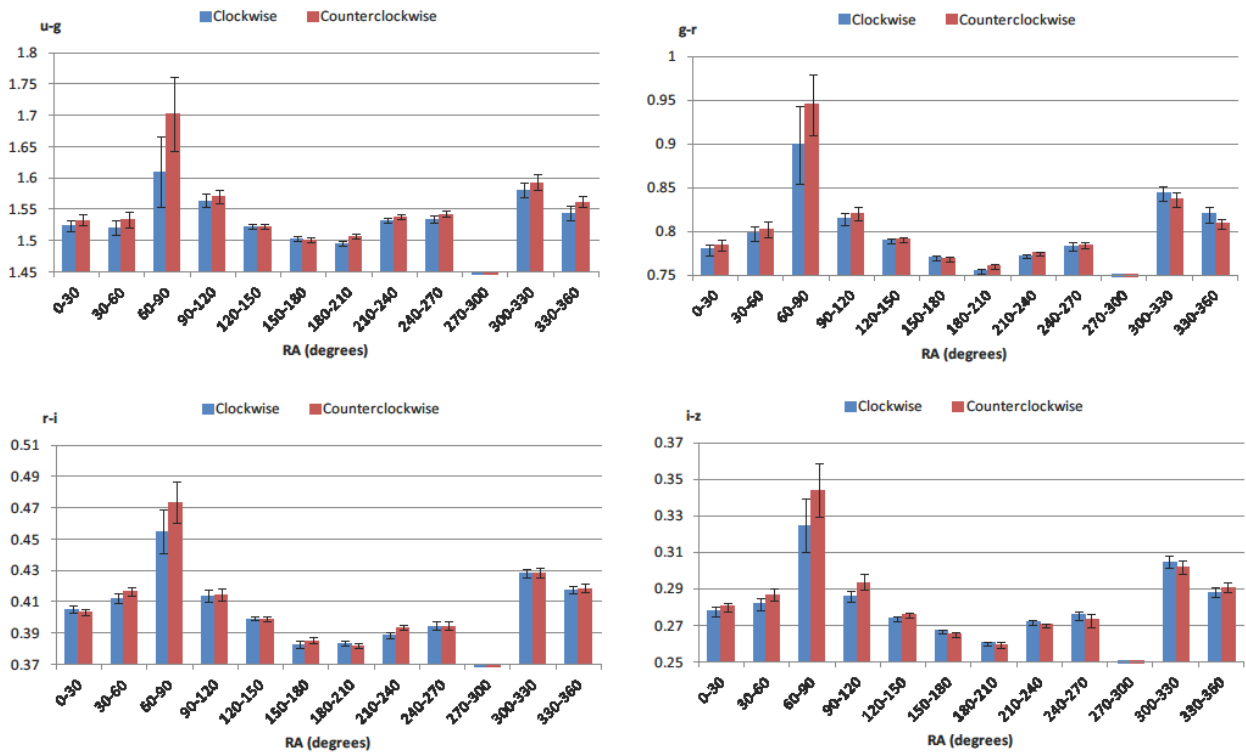

The differences in the means of u-g, g-r, r-i, i-z were also tested in different redshift ranges. Table 3 shows the t-test two-tailed probability that the means computed for galaxies with different handedness are not different. As the table shows, when separating the data into different redshift ranges the highest probability for u-g difference is in redshift range of 0.1-0.2.

**Table 3.** Two-tailed t-test probability of difference between the means in clockwise galaxies and counterclockwise galaxies in different redshift ranges.

| Color | 0<z<0.1 | 0.1<z<0.2 | 0.2<z<0.3 |
|---|---|---|---|
| u-g | ~0.0843 | ~0.0115 | ~0.969 |
| g-r | ~0.1897 | ~0.1894 | ~0.1600 |
| r-i | ~0.7056 | ~0.9653 | ~0.3832 |
| i-z | ~0.5773 | ~0.3893 | ~0.1685 |

To profile the asymmetry, the u-g color differences were computed for different RA and DEC ranges. Tables 4 and 5 show the asymmetries of galaxies with z<0.2 and 0.2<z<0.3, respectively, in three declination ranges (-11.2°-15°, 15°-45°, and 45°-70.3°) and 60 degree RA ranges.

**Table 4.** Difference between u-g of galaxies with different handedness in different RA and DEC ranges for galaxies with z<0.2. The two-tailed t-test P values are in parenthesis. Empty cells are ranges with no galaxies.

| RA | -11.2°-15° | 15°-45° | 45°-70.3° |
|---|---|---|---|
| **0°-60°** | 0.00734 (0.3968) | 0.015638 (0.6088) | |
| **60°-120°** | 0.05180 (0.2956) | 0.000311 (0.9846) | -0.02951 (0.4494) |
| **120°-180°** | 0.00655 (0.3004) | 0.007561 (0.1195) | -0.02711 (0.0140) |
| **180°-240°** | 0.02074 (0.0168) | 0.000835 (0.8560) | 0.017058 (0.0236) |
| **240°-300°** | 0.017973 (0.4604) | 0.01367 (0.1252) | -0.00766 (0.6705) |
| **300°-360°** | 0.019131 (0.0798) | 0.020672 (0.0657) | -0.01812 (0.6404) |

**Table 5.** Difference between u-g of clockwise and counterclockwise galaxies in different RA and DEC ranges for galaxies with 0.2<z<0.3. The two-tailed t-test P values are in parenthesis.

| RA | -11.2°-15° | 15°-45° | 45°-70.3° |
|---|---|---|---|
| **0°-60°** | | 0.045886 (0.9452) | |
| **60°-120°** | | -0.03738 (0.699) | 0.432131 (0.6249) |
| **120°-180°** | -0.03059 (0.442) | -0.02609 (0.3519) | -0.09046 (0.2453) |
| **180°-240°** | | -0.00244 (0.7946) | -0.02813 (0.5691) |
| **240°-300°** | 0.041818 (0.8512) | 0.033898 (0.4794) | -0.37415 (0.3684) |
| **300°-360°** | 0.01202 (0.9237) | | |

In the z range of >0.2 the number of galaxies is low, leading to high t-test P values, as well as many RA,DEC ranges in which no galaxies were found in SDSS DR7. In z<0.2 the highest asymmetry was measured between RA 180° and 240°.

Figure 2 shows the histogram of the distribution of clockwise and counterclockwise galaxies by their u-g. The numbers are normally higher for the clockwise galaxies because there are more clockwise galaxies in the dataset than counterclockwise galaxies.

**Figure 2.** Histogram of galaxies with different handedness by u-g.

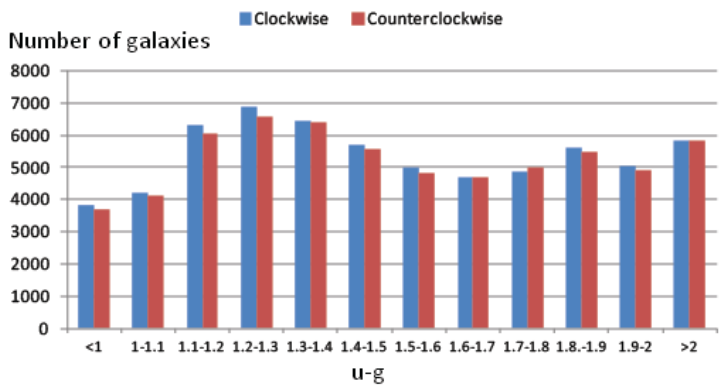

For control, the analysis was also done such that the galaxies were separated to two arbitrary different groups such that the first group included all galaxies with an even SDSS ID, and the other group included all galaxies with odd SDSS ID. The results showed no statistically significant between the color of the galaxies in the two groups. For instance, the mean u-g in the galaxies with even IDs is ~1.521188±0.0017, and the mean u-g in the galaxies with odd ID is ~1.521131±0.0017, leading to a t-test P value of 0.9811 that the two values are not different.

## 4. Conclusions

The spin direction of a spiral galaxy is an obvious and noticeable morphological feature. In this paper we described color differences between galaxies with different handedness, and showed that galaxies that spin counterclockwise have a higher u-g compared to galaxies that spin clockwise, and the difference is the most significant in redshift range of 0.1-0.2 and RA range $180^o$-$210^o$. While the statistical significance of the u-g difference is marginal, future sky surveys such as LSST will acquire more and better galaxy images that will allow morphological analysis of more galaxies and consequently stronger statistical significance of the results.

The comparisons were based on galaxies in the same region of the sky, but were separated by their handedness so that the color difference cannot be attributed to the different atmospheric effects when different parts of the sky are observed. The number of galaxies used in this study is not sufficient to profile statistically significant asymmetry in each RA range, making it difficult to profile a possible dipole axis [10,11]. Also, the fact that different regions of the sky were imaged on different days of the year adds further difficulties in deducing valid conclusions that are based on the comparison of the asymmetries in different sky regions, where differences between regions might be driven by differences in the image acquisition conditions when imaging different regions of the sky. However, while the strength of the u-g asymmetry between galaxies with different handedness can vary between different regions of the sky because of calibration or atmospheric effects, u-g differences between clockwise and counterclockwise galaxies are not expected in the same sky region. It is important to note that the galaxies used in this study are not uniformly distributed in the sky, but their distribution follows the distribution of galaxies in SDSS.

Galaxy mergers can change the spin direction of the galaxy while also increasing star formation [13], and therefore large-scale blue color asymmetry can also be related to large-scale asymmetry in interaction between galaxies, or to large-scale asymmetry in spin direction [10,11,14]. Another possible direction of explanation can be internal extinction at the edges of the arms, which can be affected by the radial velocity. While there is no solid explanation to the surprising difference between the colors of galaxies based on their spin rotation, the purpose of this study is to initiate a discussion and further investigation on difference between galaxies based on their spin direction.

**Acknowledgments**

I would like to thank the three anonymous reviewers for their insightful comments and the thorough discussion that helped to improve the paper. I would also like to thank John Wallin for his help in obtaining the spectroscopy data for the Galaxy Zoo galaxies.

**Conflicts of Interest**

The author declares no conflict of interest.